# Software Effort Estimation Based on Optimized Model Tree

Mohammad Azzeh
Faculty of Information Technology
Applied Science University
Amman, Jordan POBOX 166
m.y.azzeh@asu.edu.jo

## ABSTRACT

**Background**: It is widely recognized that software effort estimation is a regression problem. Model Tree (MT) is one of the Machine Learning based regression techniques that is useful for software effort estimation, but as other machine learning algorithms, the MT has a large space of configuration and requires to carefully setting its parameters. The choice of such parameters is a dataset dependent so no general guideline can govern this process which forms the motivation of this work. **Aims**: This study investigates the effect of using the most recent optimization algorithm called Bees algorithm to specify the optimal choice of MT parameters that fit a dataset and therefore improve prediction accuracy. **Method**: We used MT with optimal parameters identified by the Bees algorithm to construct software effort estimation model. The model has been validated over eight datasets come from two main sources: PROMISE and ISBSG. Also we used 3-Fold cross validation to empirically assess the prediction accuracies of different estimation models. As benchmark, results are also compared to those obtained with Stepwise Regression Case-Based Reasoning and Multi-Layer Perceptron. **Results**: The results obtained from combination of MT and Bees algorithm are encouraging and outperforms other well-known estimation methods applied on employed datasets. They are also interesting enough to suggest the effectiveness of MT among the techniques that are suitable for effort estimation. **Conclusions**: The use of the Bees algorithm enabled us to automatically find optimal MT parameters required to construct effort estimation models that fit each individual dataset. Also it provided a significant improvement on prediction accuracy.

## Categories and Subject Descriptors

D.2.9 [Software Engineering]: Management—cost estimation.

## General Terms

Management, Measurement

## Keywords

Software Effort Estimation, Model Tree, Bees Algorithm.
.

## 1. INTRODUCTION

Estimating the likely software project effort is one of the major challenges in software engineering and has achieved a considerable interest within scientific research community [2, 3, 15, 16]. In literature, a variety of software effort estimation models have been proposed so far but they have suffered from common problems such as very large performance deviations as well as being highly dataset dependent [10]. The evaluation and comparison results of those models are often contradictory so no single model can outperform others [11, 12]. The main principal reason behind that is the nature of software datasets which are characteristically noisy. However, Software effort estimation is recognized as a regression problem [1], and machine learning methods such as Regression Tree [9], Model Tree (MT) [23], Support Vector Machine [1], Radial Basis Functions, etc. are more capable of handling noisy datasets than statistical based regression models that focus on the correlation between variables. The main concern of this paper will focus on MT. MT [23, 25] is a special type of decision tree and regression tree, but unlike regression tree that have numerical values at the leaves, the MT have linear functions as illustrated in Figure 1. MT is one of the powerful methods for performing regression since it can include categorical features in constructing such model without the need to convert them into dummy variables as performed in the basic regression models. But, like other machine learning techniques, the performance of MT is a data dependent and has large space of configuration possibilities and design options induced for each individual dataset. So it is not surprise to see contradictory results and different performance figures when make slight changes to MT parameters. Such parameters include selection of minimum number of cases ($C$) that one node may represent, Whether to prune the tree ($P$), finding smoothing coefficient ($K$), and split threshold($T$).

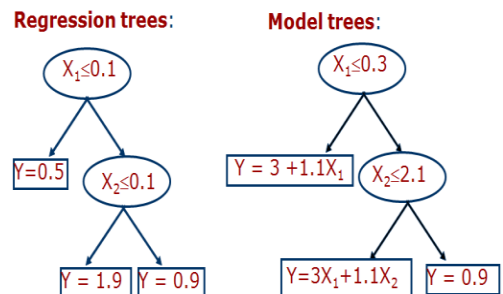

Figure 1. Difference between regression trees and Model Trees.

Since the selection of these parameter values is subjective and dataset dependent, there is no general guideline can govern this process which forms the motivation of this work. In this paper we employed Bees algorithm [18] to search for the optimal design options of MT that fit a specific dataset. The Bees algorithm is a new population-based search algorithm, it was first proposed by Pham [18]. The algorithm mimics the food foraging behavior of swarms of honey bees. In its basic version, the algorithm performs a kind of neighborhood search combined with random search and can be used for optimization. The present paper investigates the effect on the improvement of effort estimation accuracy in MT when the Bees algorithm method is adopted to optimize MT parameters. To the best of our knowledge the Bees algorithm has not been used in software effort estimation and this paper shows a potential to provide more accurate results.

The rest of this paper is organized as follows: Section 2 presents background to effort estimation. Section 3 presents methodology of this study and the proposed approach. Section 4 presents design of experiments for this study. Section 5 presents threats to study validity. Section 6 presents results of empirical validation. Finally, section 7 summarizes our work and outlines the future studies.

## 2. BACKGROUND

Software effort estimation has been an active research topic in software engineering for more than four decades [2, 5, 8, 21]. The advances in data mining and machine learning encouraged researchers to start investigating techniques previously used within other fields such as Fuzzy systems [3, 7], Regression tree [9], Evolutionary computing [6] Case based reasoning [13, 14, 20, 22], Support vector regression [1] Rule based expert systems [19] and neural networks [12]. These alternatives have two distinct advantages: (1) the capability to model complex set of relationships between dependent variable and the independent variables. (2) The capability to learn from historical project data. However, Software effort estimation is a regression problem [17] and the machine learning based regression are appropriate for this kind of problem. In this paper we focus on MT. Although quite emphasis was placed on neural network based regression [ ], support vector regression [1] and stepwise regression [14] there is quite small research papers investigated the performance of MT in software effort estimation such as [1, 17]. These studies reported that the choice of parameters values have strong impact on the accuracy of MT where inappropriate setting can lead to over/under fitting, that is, bad estimation. In other words, the relative performance of MT depends on the size and the characteristics of the dataset. This finding has motivated the current research work.

The MT was first proposed by Quinlan [23, 25] and is considered a special type of decision tree model developed for the task of regression. However, the main difference between MT and basic regression trees is that the leaves of regression trees present numerical values only, whereas the leaves of a MT have linear functions as illustrated in Figure 1. The general model tree building methodology allows input variables to be a mixture of continuous and categorical variables which is considered useful for software datasets that have complex structure. The final MT consists of a tree with linear regression functions at the leaves, and the prediction for an instance is obtained by sorting it down to a leaf and using the prediction of the linear model associated with that leaf. The principle behind MTs is fairly simple, that is, it is constructed through a process known as binary recursive partitioning method. This is an iterative process of splitting the data into partitions, and then splitting it up further on each of the branches [25]. A tree is formed of decision nodes where a test is made against data for a given instance, and a choice must be made about which of the node's children is traversed next. The choice of each node of the tree is usually guided by a least squares error criterion. The real strength of MTs, however, lies in their inherent simplicity, and the ease with which they can be interpreted by non-experts in either computing or the particular application subject.

In this paper the M5P algorithm [23, 25] has been adopted to develop MT based software effort estimation. M5P is a powerful implementation of Quinlan's M5 algorithm for inducing both Model Trees and Regression Trees [23]. Each training sample has a set of attributes that can be either numeric or categorical. The objective of M5P is to construct a model using the training set that is able to predict continuous dependent variable. The procedure of Model tree construction using M5P algorithm consists of three main steps [25]:

1. <u>Tree construction</u>: In the first phase, a standard regression tree is grown whereby the decision-tree induction algorithm is used to construct MT by minimizing the intra-subset variation in the class values down each branch in order to split the tree. The 'purity' measure for the splitting criterion is the standard deviation of the class values of the examples at a node: the algorithm selects the attribute to split on as the one giving the largest decrease in standard deviation. The splitting procedure can stop when only few instances remain or when the class values of all instances that reach a node vary very slightly. After the initial tree is grown, a linear regression model is built for every node in the tree.

2. <u>Pruning</u>: The constructed tree from step 1 is then pruned back from each leaf such that each inner node becomes a leaf node with a regression plane. The main objective of pruning is to find an estimate of the 'true error' for the subtree and the regression function at every node of the tree. The true error is the expected error of the subtree/regression on unseen instances that are sorted down to that node in the tree. The outcome of this stage is the pruned tree.

3. <u>Smoothing</u>: a smoothing procedure is performed to avoid sharp discontinuities between adjacent linear models at the leaves of the pruned tree (i.e. the target value could change considerably if the value for an attribute varies a bit so that it is sorted into a different leaf). This procedure combines the leaf model prediction with each node along the path back to the root, smoothing it at each of these nodes by combining it with the value predicted by the linear model for that node.

In order to construct optimized MT (OMT), several parameters must be initially set. These parameters vary from dataset to another depending on the characteristics of that dataset. Practically, it is very hard to discover appropriate values for such parameters unless an optimization algorithm is applied. However, these parameters are:

- The minimum number of training data cases ($C$) one node may represent.

- Prune (P): Whether to prune the tree.
- Smoothing coefficient (*K*) for the smoothing process. For larger values, more smoothing is applied. For large (relatively to the number of training data cases) values, the tree will essentially behave like containing just one leaf (corresponding to the root node). For value 0, no smoothing is applied.
- Splitting Threshold (*T*): Where the splitting procedure should stop. A node is not splitted if the standard deviation of the output variable values at the node is less than Splitting Threshold of the standard deviation of the output variable values of the entire original data set.

## 3. METHODOLOGY

The proposed estimation method exploits the search capability of the Bees algorithm to overcome the local optimum problem of the MT. The Bees Algorithm [18] is smart optimization algorithm inspired by the natural foraging behavior of honey bees to find the optimal solution. The algorithm performs a kind of neighborhood search combined with random search and can be used for both combinatorial optimization and functional optimization. The proposed Model Tree technique exploits the search capability of the Bees Algorithm to overcome the local optimum problem of the parameter tuning. More specifically, the task is to search for appropriate parameter values such that the performance measure (*MMRE*) is minimized.

Before starting, the BA parameters must be carefully set [18], these parameters are: problem size (*Q*), number of scout bees (*n*), number of sites selected out of *n* visited sites (*s*), number of best sites out of *s* selected sites (*e*), number of bees recruited for best *e* sites (*nep*), number of bees recruited for the other selected sites (*nsp*), other bees number (*osp*) and initial size of patches (*ngh*) which includes site and its neighborhood in addition to Stopping criterion which in our study is to minimize *MMRE* performance measure.

The algorithm starts with an initial population of *n* scout bees. Each bee represents a potential solution as set of four MT parameter values (*C*, *P*, *K*, *T*). The scout bees are placed randomly in initial search space to visit enormous number of sites (solutions). The fitness computation process is carried out for each site visited by a scout bee by calculating *MMRE*. This step is essential for colony communication which shows the direction in which flower patch will be found, its distance from the hive and its fitness [18]. This information helps the colony to send its bees to flower patches precisely, without using guides or maps. Then, the best sites visited by the highest fittest bees are being selected for neighborhood search. The area of neighborhood search is determined by identifying the radius of search area from best site which is considered the key operation of Bees algorithm. The algorithm continues searching in the neighborhood of the selected sites, recruiting more bees to search near to the best sites which may have promising solutions. The bees can be chosen directly according to the fitnesses associated with the sites they are visiting. Alternatively, the fitness values are used to determine the probability of the bees being selected. Then the fittest bee from each patch is selected to form the next bee population. The claim here is to reduce the number of points to be explored. Finally the remaining bees are assigned to search randomly for new potential solutions. These steps are repeated until the criterion of stop is met or the number of iteration has finished. At the end of each iteration, the colony of bees will have two parts to its new population – those that were the fittest representatives from a patch and those that have been sent out randomly. The **pseudo code** of Bees algorithm is shown in Figure 2.

```
Input:Q,n,s,e,ngh,nep,nsp
Output:BestBee
Population ← InitializePopulation(n,Q)
While(!StopCondition())
    MMRE=EvaluatePopulation(Population,MT)
    BestBee ← GetBestSolution(Population)

    NextGeneration ← ø
    ngh ← (ngh × PatchDecreasefactor)
    Sites_best ← SelectBestSites(Population , s)
    for(Site_i∈ Sites_best)
        nsp ← ø
        if(i<e)  nsp ← nep
        else  nsp ← osp
        Neighborhood ← ø
        for(j To nsp)

    Neighborhood ← CreateNeighborhoodBee(Site_i,ngh)
        endfor
        NextGeneration ← GetBestSolution(Neighborhood
)
    endfor
    RemainingBeesnum ← (n-s)
    for(j To RemainingBeesnum)
        NextGeneration ← CreateRandomBee()
    endfor
    Population ← NextGeneration()
endwhile
Return(BestBee)
```

**Fig. 2.** Pseudo code of OMT using Bees algorithm

## 4. EXPERIMENTAL DESIGN

Throughout this investigation we used MATLAB 7.1 to implement the optimized MT using Bees algorithm. However, the proposed approach has been evaluated using eight different datasets that exhibit different characteristics. These datasets come from two different sources namely PROMISE [4] and ISBSG [26]. PROMISE is an on-line publically available data repository and it consists of datasets donated by various researchers around the world. The datasets come from this source are: Desharnais, Kemerer, Albrecht, COCOMO'81, Maxwell, Telecom and NASA93 datasets. The other dataset comes from ISBSG data repository (release 10) which is a large data repository consists more than 4000 projects collected from different types of projects around the world. Since many projects have missing values only 500 projects with quality rating "A" are considered. 14 useful attributes were selected, 8 of which are numerical attributes and 6 of which are categorical attributes. The descriptive statistics of such datasets are summarized in Table 1. From the table, we can observe that all the datasets have positive skewness values which range from 1.78 to 4.36. This observation indicates that the datasets are extremely heterogeneous. As consequence we make sure that we test the proposed model adequately.

For each dataset we follow the same testing strategy, we used 3-Fold cross validation to identify test and train projects such that, in each run, we select one set as test set and the remaining sets as training set. This procedure is performed until all projects within dataset are used as test projects. The main reason for using 3-fold cross validation is that, this procedure has been widely used to evaluate machine learning based effort estimation methods such as [14]. In each run, The prediction accuracy of different techniques is assessed using MMRE, PRED(0.25) performance measure. *MMRE* computes mean of the absolute percentage of error between actual and predicted project effort values as shown in Eq. 1. PRED(0.25) is used as a complementary criterion to count the percentage of estimates that fall within less than 0.25 of the actual values. We also used median of MREs (MdMRE).

$$MMRE = \sum_{i=1}^{N} \frac{|Effort(p_i) - \overline{Effort(p_i)}|}{Effort(p_i)} \quad (1)$$

Where $Effort(p_i)$ and $\overline{Effort(p_i)}$ are the actual value and predicted values of project $p_i$.

$$PRED(0.25) = \frac{\lambda}{N} \times 100 \quad (2)$$

Where $\lambda$ is the number of projects that have magnitude relative error less than 0.25, and $N$ is the number of all observations.

We also used Boxplot of absolute residuals as alternatives to simple summary measures because they can give a good indication of the distribution of residuals and can help explain summary statistics such as MMRE and PRED(0.25). On the other hand, we used Wilcoxon sum rank test to investigate the statistical significance of all the results, setting the confidence limit at 0.05. The Wilcoxon sum rank test is a nonparametric test that compares the medians of two samples. The reason behind using these tests is because all absolute residuals for all models used in this study were not normally distributed.

In turn, the obtained results from the proposed approach have benchmarked to other frequently used prediction techniques such as Case-Based Reasoning (CBR) [20], Multilayer Perceptron Neural Network (MLP) and stepwise regression (SWR). In this study we used the original CBR model that was proposed by Shepperd et al. [20] which uses Euclidian distance as similarity measure, No feature selection, using only one analogy without adjustment. On the other hand, before developing Stepwise regression model we should make sure that assumptions related to using stepwise regression are not violated [14]. For example, skewed numerical variables need to be transformed such that they resemble more closely a normal distribution. The logarithmic transformation ensures that the resulting model goes through the origin on the raw data scale. It also caters for both linear and non-linear relationships between size and effort. The ANN work was based on a simple multi-layer perceptron with a back propagation learning algorithm using the software tool MATLAB. In configuring the network we had to make design decisions concerning the topology, learning rate and momentum. Each configuration was also tested ten times to assess the impact of different initial random weights for the nodes.

Table 1 Statistical properties of the datasets

| Dataset | Cases # | min | max | Effort mean | skewness |
|---|---|---|---|---|---|
| ISBSG | 500 | 668 | 14938 | 2828.5 | 2.09 |
| Desharnais | 77 | 546 | 23940 | 5046.3 | 1.96 |
| COCOMO | 63 | 5.9 | 11400 | 683.5 | 4.36 |
| Kemerer | 15 | 23.2 | 1107.3 | 219.2 | 2.76 |
| Albrecht | 24 | 0.5 | 105.2 | 21.87 | 2.15 |
| Maxwell | 62 | 583 | 63694 | 8223.2 | 3.27 |
| NASA93 | 18 | 8.4 | 824 | 624.4 | 4.18 |
| Telecom | 18 | 23.45 | 1115.5 | 284.3 | 1.78 |

## 5. THREATS TO VALIDITY

This section presents the comments on the validities of our study based on the internal and external threats to validity. In our opinion the greatest threats are to the internal validity of this study; i.e. the degree to which conclusions can be drawn with regard to the better parameter setup for MT based effort prediction. One possible threat to internal validity is the chosen of The Bees algorithm parameters. Currently, there is no gaudiness for how to determine such parameters for every single dataset. Although it is well recognized that use of appropriate parameters has a strong impact on the identification of optimal solutions, in this study we determined the parameter values of Bees algorithm based on empirical investigation. However, we believe this decision was reasonable even though its intensive computation cost. Despite special emphasis was placed on the effectiveness of the performance measure used as stopping criterion, complete certainty with regard to this issue was challenged and we had to rely on common performance measure (MMRE) which we no longer believe to be a completely trustworthy accuracy indicator. We do not consider that choice was a problem since this study was motivated with the finding from previous studies that used MMRE as optimization criterion [6]. In order to make apple-to-apple comparisons between different adaptation techniques we preferred to use 3-Fold cross validation strategy, though some authors favored one leave out cross validation. The principal reason is that, the 3-Fold cross validation has been used in some previous studies and recommended by [14] to do comparison between different estimation models. With regard to external validity, i.e. the ability to generalize the obtained findings of our comparative studies, we consider that some datasets are very old to be used in software cost estimation because they represent different software development approaches and technologies. The reason to this is that these datasets are publically available, and still widely used for comparison purposes.

## 6. RESULTS

Prior to conducting the experiments, it was necessary to decide what parameter values of Bees algorithm to adopt. This part has proved very difficult to investigate thoroughly which is unfortunate since we believe to be of considerable significance. The difficulty is that we have no gaudiness and efficient means of finding the optimal parameter values for each single datasets. Therefore the decision was made to using sufficient number of scout Bees and elite Bees to investigate thoroughly all potential sites (solutions). Specifically, we conducted several empirical studies on employed datasets to investigate the most appropriate parameter values that can provide good results. This resulted in the recommended values shown in Table 2. The second decision was to specify how to find neighborhoods of the best visited sites within allowed search area (i.e. given radius). Here we developed a small piece of code to automatically find potential distance between best visited site and its potential neighborhood taken into account the restriction on the radius of search area. However, its performance cannot be assessed individually until it is combined with the whole algorithm. The algorithm is run until it converges, i.e. until the change in MMRE is smaller than some small constant, or to a maximum of 50 optimization iterations.

Table 2 Parameter values of Bees algorithm

| Parameter | Value |
|---|---|
| Q | 4 values ($C, P, K, T$) |
| N | 30 |
| S | 30 |
| E | 20 |
| Nep | 15 |
| Nsp | 30 |
| Osp | 20 |
| Ngh | 15 |
| **Stopping criterion** | Minimize MMRE |

For the experiments we used the 8 benchmark datasets from the PROMISE and ISBSG repository given in Table 1. Their size ranges from a few tens to a few hundred projects, they contain varying numbers of numeric and nominal attributes and some contain missing values. For every dataset, we performed ten runs of three fold stratified cross-validation (using the same splits into training/test set for every method). The results shown in Table 3 summarize the relative testing and training accuracies of the optimized MT using the MMRE, MdMRE and PRED values for all employed data sets. In short, the obtained results for all data sets are promising as being predictive especially in terms of MMRE. Although the performance figures on Maxwell data set was poor, it is still considered promising if we take into account the structure of Maxwell that contains too many categorical features. The notable results from this table are for NASA93, Albrecht, Desharnais, and COCOMO data sets where the optimized MT obtained good results. We can also observe that the performance of optimized MT over Kemerer and NASA93 was notable in terms of MdMRE which would suggest the suitability of this model for small size datasets. Early results from using Bees algorithm, suggest that optimizing one particular measure, in this case MMRE has the effect of degrading a lot of other measures and the fitness function that is not specifically tied to one particular measure may give acceptable over all accuracy. This is illustrated in Table 3 which shows superior results for MMRE but poor results of PRED. This would suggest that more research is required not only on Bees algorithm, but also on the problem itself as to which of the many performance measures or combinations of measures is the more appropriate in practice.

Table 3 Prediction accuracy of optimized MT

| Dataset | MMRE | | MdMRE | | PRED% | |
|---|---|---|---|---|---|---|
| | Test | Train | Test | Train | Test | Train |
| ISBSG | 49.0 | 42.0 | 42.7 | 37.8 | 29.1 | 34.6 |
| Albrecht | 26.6 | 28.2 | 22.6 | 19.7 | 50.0 | 50.0 |
| Kemerer | 37.4 | 30.4 | 15.8 | 15.0 | 73.3 | 76.6 |
| Desharnais | 32.3 | 41.5 | 26.8 | 28.2 | 45.5 | 52.0 |
| COCOMO | 32.8 | 37.3 | 22.4 | 23.3 | 58.3 | 54.2 |
| Maxwell | 51.6 | 44.6 | 43.2 | 37.2 | 33.9 | 40.0 |
| Telecom | 34.7 | 27.2 | 24.2 | 22.3 | 50.0 | 55.0 |
| NASA93 | 14.6 | 16.2 | 6.1 | 8.1 | 83.3 | 81.0 |

Concerning number of features appearing at the leaves of every training dataset, it was observed that not all features have been selected. Although the fraction of features that are discarded varies wildly, in most cases the number of features included in the final model is reduced substantially. On average, the biggest reduction takes place for datasets with a high number of attributes that are not too large.

Table 4 shows the result of comparing OMT to CBR, SWR and MLP. In all datasets the OMT method had the minimum MMRE (i.e. the best performance among all investigated methods) whereas MLP had the poorest performance among all variants. However, these findings are indicative of the performance of using Bees algorithm to find optimal choice of MT parameter values which lead to significant improvement on the overall accuracy. Moreover, we can also observe that the performance of SWR is better than CBR over six out of eight datasets which also confirm that the software effort estimation is a regression problem and regression based technique is more capable for its datasets. Interestingly, unlike SWR the results of OMT were obtained without making feature pruning and all features have been used so it is a good point for OMT. Nevertheless, this highlights an issue that can the performance of OMT be improved when applying feature subset selection algorithms. In addition to the performance of OMT over datasets that contain categorical features, it was observed that OMT can deliver reasonable estimates and better than obtained by other methods for the datasets that contain only numerical features (e.g. Albrecht, NASA93, and Telecom). This is because the OMT uses decision tree to classify feature values based on their standard deviation and construct initial tree which then is pruned back to reduce estimate errors.

Figures 3 to 10 show Boxplot of absolute residuals of different models over the employed datasets. These figures show number of interesting findings summarized in the following points:

1. ISBSG, Albrecht, Kemerer, Desharnais and NASA93 datasets (Figures 3 to 6 and 10): median and box length of absolute residuals of OMT is smaller than others which demonstrate reduced variability of absolute residuals and confirm that OMT is better than other methods.
2. COCOMO, Maxwell and Telecom dataset (Figures 7 to 9): OMT and SWR are at the same level of accuracy because they have relatively similar median of absolute residuals and smaller box length. Also the use of OMT produced less outliers so the use of OMT is more reliable.

The absolute residuals of all methods are taken and compared using Wilcoxon sum rank test ($\alpha=0.05$) and the results are presented in Table 5. The test reports 16 significant differences in favor of OMT. Note that all wins are not only for larger datasets but also on the smaller datasets. This is not surprising, because overfitting is generally more of a problem when only few training examples are available such as in MLP, and the use of OMT to identify best features for linear regression at leaves helps to prevent overfitting. Predictions based on OMT model presented statistically significant but necessarily accurate estimations than MLP over all datasets except COCOMO dataset. This would suggest that there is a significant difference if the prediction generated by OMT and MLP. The results of OMT were also significant than CBR over five datasets, and SWR over four datasets. For ISBSG, Desharnais and NASA93 we can notice that OMT produced significant results than original other methods. Surprisingly, the statistical test results over Albrecht and Telecom demonstrate that there is no significant difference if the predictions generated by OMT or CBR and SWR. In other words, although there is some evidence that accuracy improves using OMT we cannot show statistical significance ($\alpha=0.05$). Note that in all cases we generate more accurate predictions than merely using the CBR and SWR, however, we are unable to demonstrate statistical significance. This suggests that a certain degree of caution should be exercised when comparing prediction systems over small datasets even if there appear to be difference in accuracy indicators such as MMRE and PRED.

In this paper we have looked at the some of the reported experiences of using MT technology for software effort estimation. In particular, we have considered the question of how can we optimize MT based effort estimation to adapt and fit every single dataset. We have argued that these arise in part due to different characteristics of the datasets being examined. We also argued, and this is the major theme of our experimentation, that tuning MT based estimation technique is a non-trivial task. We focus, however, upon using Bees algorithm for selecting appropriate MT parameter values for each employed dataset. Our analysis suggests that decisions on how to configure the MT technique can have a substantial impact upon the level of accuracy obtained. For example, with the NASA93 dataset we obtained the best performance ever had over this dataset. From the empirical analysis it can be observed that, at least for the sample of data sets investigated, OMT is superior in terms of prediction accuracy and statistical significance results, to estimation based on CBR, SWR and MLP. Although it must be remembered that the OMT can be optimised on the appropriate choose of parameter values, the results obtained have shown that the use of Bees algorithm to find optimal parameter values of MT has strong impact on the prediction accuracy of MT. Further, from the results, OMT would seem to show number of interesting advantages. First, OMT remains viable when using too many categorical features (e. g. the Maxwell dataset). Second, OMT remains accurate for small data sets (e. g. the NASA93, Telecom, Albrecht and Kemerer), as well as for large datasets (e.g. ISBSG). Third, OMT remains accurate where the number of features is limited (e. g. the Telecom and NASA93 dataset2). The main disadvantage for MT is that when developing MT the possible values of categorical features are detected from the training data so any new values detected later (e.g. in the test data) will be treated as null. Nevertheless, the MT still has the capability to discover the suitable path for each test data using other continues variables.

## 7. CONCLUSIONS

This paper presents a new approach for improving software effort estimation accuracy based on the use of optimized Model Tree. We used the Bees algorithm to search for the optimal parameter values of MT that adapt each single dataset. The use of the Bees algorithm enabled us to obtain an automatic choice of the parameters required to run MT, and a significant improvement on prediction accuracy for MT based effort estimation. One of the advantages of the proposed method is that it does not become trapped at locally optimal solutions. This is due to the ability of the Bees algorithm to perform local and global search simultaneously. While we are not guaranteed that the obtained performance figures are the global optimum, the results we are presenting are the best performance when using all features without pruning. Overall, we are encouraged by the results of the present study, which although minor in their own right, they give much more strongly to the value of MT for effort estimation when combined with other results. Nevertheless, Publication of raw results is still important so further research is necessary to investigate whether the use feature subset selection can also help in obtaining accurate estimates.

## 8. ACKNOWLEDGMENTS

The author is grateful to the Applied Science University, Amman, Jordan for financial support granted to cover the publication fee of this research article.

**Table 4** Prediction accuracy of optimized MT

| Dataset | OMT | | CBR | | SWR | | MLP | |
|---|---|---|---|---|---|---|---|---|
| | MMRE | PRED% | MMRE | PRED% | MMRE | PRED% | MMRE | PRED% |
| ISBSG | **49** | 29.1 | 86.8 | **31.8** | 88.3 | 13.5 | 103.0 | 17.6 |
| Albrecht | **26.56** | **50.0** | 66.8 | 37.5 | 47.0 | 33.3 | 86.6 | 25.0 |
| Kemerer | **37.4** | **73.3** | 64.0 | 40 | 101 | 20.0 | 101.0 | 13.3 |
| Desharnais | **32.3** | **45.5** | 77.7 | 28.6 | 62.9 | 36.4 | 152.7 | 18.2 |
| COCOMO | **32.8** | **58.3** | 166.8 | 10 | 36.1 | 51.7 | 167.0 | 33.3 |
| Maxwell | **51.6** | **33.9** | 166.0 | 14.5 | 57.2 | 21.0 | 174.7 | 22.6 |
| Telecom | **34.7** | 50.0 | 125.5 | 33.3 | 42.73 | **61.1** | 203.1 | 22.2 |
| NASA93 | **14.6** | **83.3** | 53.4 | 33.3 | 24.6 | 66.7 | 79.5 | 33.3 |

**Table 5.** Wilcoxon sum rank test

| Dataset | CBR | SWR | MLP |
|---|---|---|---|
| ISBSG | **<0.01** | **<0.01** | **<0.01** |
| Albrecht | 0.15 | 0.11 | **0.04** |
| Kemerer | 0.16 | **<0.01** | **0.01** |
| Desharnais | **<0.01** | **0.04** | **<0.01** |
| COCOMO | **<0.01** | 0.76 | 0.19 |
| Maxwell | **<0.01** | 0.47 | **0.03** |
| Telecom | 0.26 | 0.76 | **0.02** |

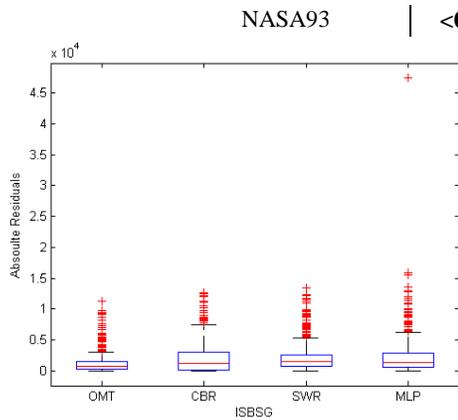
Figure 3. Boxplot of ISBSG dataset

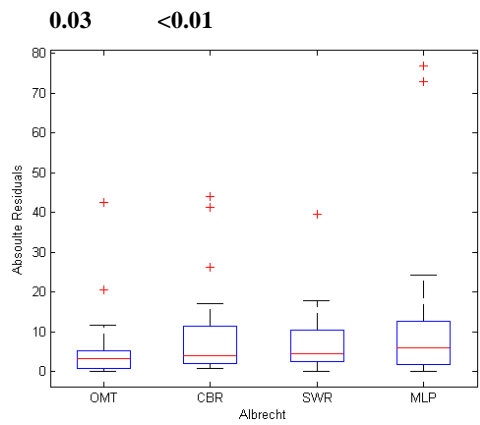
Figure 4. Boxplot of Albrecht dataset

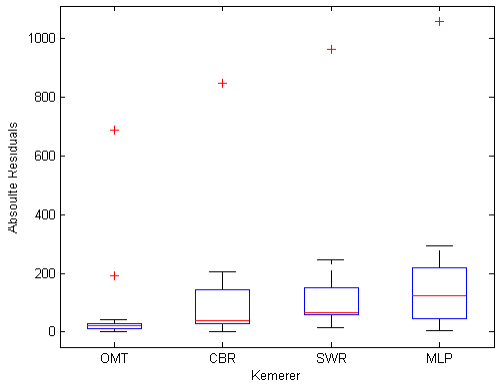
Figure 5. Boxplot of Kemerer dataset

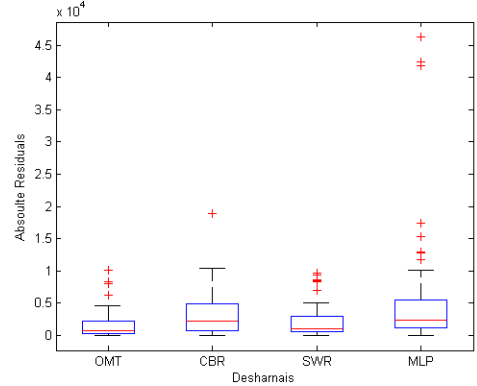
Figure 6. Boxplot of Desharnais dataset

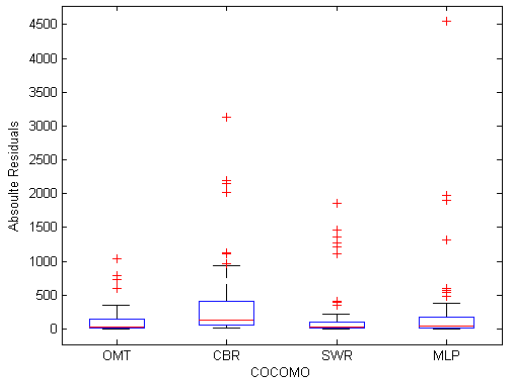
Figure 7. Boxplot of COCOMO dataset

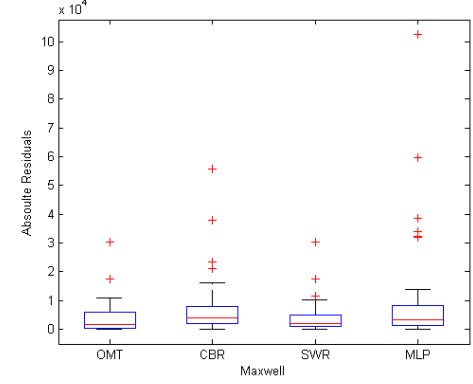
Figure 8. Boxplot of Maxwell dataset

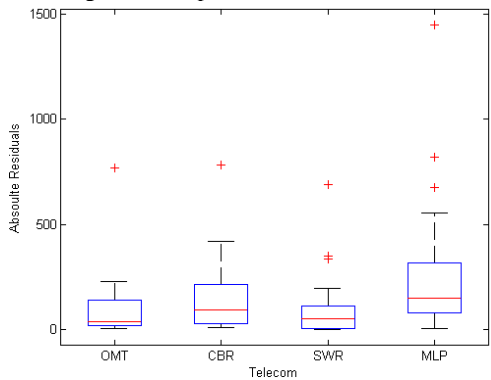
Figure 9. Boxplot of Telecom dataset

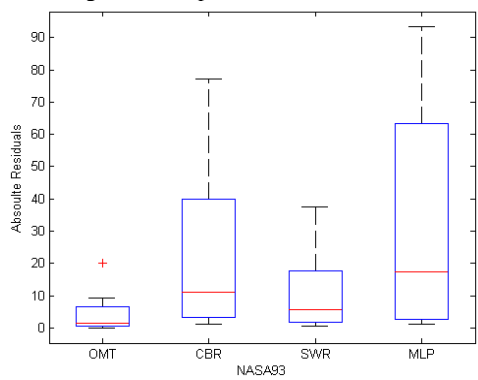
Figure 10. Boxplot of NASA93 dataset